\documentclass[fleqn,usenatbib]{mnras}
\usepackage{newtxtext,newtxmath}

\usepackage[T1]{fontenc}

\DeclareRobustCommand{\VAN}[3]{#2}
\let\VANthebibliography\thebibliography
\def\thebibliography{\DeclareRobustCommand{\VAN}[3]{##3}\VANthebibliography}

\usepackage{graphicx}	
\usepackage{color,xcolor}
\definecolor{orange}{HTML}{FF7F00}

\title[Debiasing Standard Siren Cosmology]{Debiasing Standard Siren Inference of the Hubble Constant with Marginal Neural Ratio Estimation}

\author[Gagnon-Hartman et al.]{
Samuel Gagnon-Hartman,$^{1,2,3}$\thanks{samuel.gagnonhartman@sns.it}
John Ruan$^{1}$, Daryl Haggard$^{2}$
\\
$^{1}$Department of Physics and Astronomy, Bishop's University, 2600 College Street, Sherbrooke J1M 1Z7, Canada \\
$^{2}$Department of Physics and McGill Space Institute, McGill University, Montreal, QC, Canada H3A 2T8 \\
$^{3}$Scuola Normale Superiore, Piazza dei Cavalieri 7, 56126 Pisa, Italy}

\date{Accepted XXX. Received YYY; in original form ZZZ}

\pubyear{2022}

\begin{document}
\label{firstpage}
\pagerange{\pageref{firstpage}--\pageref{lastpage}}
\maketitle

\begin{abstract}
Gravitational wave (GW) standard sirens may resolve the Hubble tension, provided that standard siren inference of $H_0$ is free from systematic biases. However, standard sirens from binary neutron star (BNS) mergers suffer from two sources of systematic bias, one arising from the anisotropy of GW emission, and the other from the anisotropy of electromagnetic (EM) emission from the kilonova. For an observed sample of BNS mergers, the traditional Bayesian approach to debiasing involves the direct computation of the detection likelihood. This is infeasible for large samples of detected BNS merger due to the high dimensionality of the parameter space governing merger detection. In this study, we bypass this computation by fitting the Hubble constant to forward simulations of the observed GW and EM data under a simulation-based inference (SBI) framework using marginal neural ratio estimation. A key innovation of our method is the inclusion of BNS mergers which were only detected in GW, which allows for estimation of the bias introduced by EM anisotropy. Our method corrects for $\sim$90$\%$ of the bias in the inferred value of $H_0$ when telescope follow-up observations of BNS mergers have extensive tiling of the merger localization region, using known telescope sensitivities and assuming a model of kilonova emission. Our SBI-based method thus enables a debiased inference of the Hubble constant of BNS mergers, including both mergers with detected EM counterparts and those without. 
\end{abstract}




\section{Introduction}

The value of the Hubble constant, $H_0$, is currently the subject of dispute as a $\sim$5$\sigma$ tension exists between the latest late-time measurement using the cosmic distance ladder by the SH0ES Team \citep{tension0} and the early-time value inferred from cosmic microwave background (CMB) anisotropies by the Planck satellite \citep{tension4}. Gravitational wave (GW) standard sirens provide an independent way to measure $H_0$, and thus have the potential to resolve this dispute \citep{gw170817}.

GW observations of binary neutron star (BNS) mergers provide an estimate of the luminosity distance ($D_L$) of each merger through modeling of its GW waveform. If the electromagnetic (EM) emission from the kilonova counterparts of the mergers are also detected, then the BNS mergers can be precisely localized, thus providing their cosmological redshifts through the host galaxy spectrum. Given a sample of BNS mergers with known redshifts and luminosity distances, $H_0$ can be inferred. This approach is known as the standard siren method of inferring the Hubble constant \citep{schutz_1986,Holz_2005}. The binary neutron star merger GW170817 was the first standard siren, providing a $\sim$10$\%$ measurement of $H_0$ \citep{gw170817a,gw170817grb,soares-santos_2017,gw170817,Hotokezaka_2019}. Forecasts predict that a 2\% measurement of $H_0$ can be achieved by combining a future sample of $\sim$50 standard sirens \citep[e.g.,][]{dalal_2006,nissanke_2010,nisanke_2013,chen2018,feeney_2019}.

In order for a standard siren measurement to resolve the tension in $H_0$, it must be free of systematic biases. Here, we seek to address two major sources of bias in standard sirens: GW anisotropy bias and EM anisotropy bias. The bias from anisotropic GW emission has been shown to inflate the value of $H_0$ inferred from standard sirens \citep{malmquist_theory}, and a method to mitigate these biases was presented by \cite{gerardi2021} using a simulation-based inference (SBI) approach. An additional source of bias is introduced by observational selection effects owing to the anisotropic EM emission from the kilonova, as detailed in \cite{chen2020}, and this additional bias was not addressed by \cite{gerardi2021}. In both sources of bias, the anisotropy of BNS merger emission produces a selection effect where mergers consistent with a high value of $H_0$ are preferentially observed. 

Correcting for bias introduced by a selection effect requires a careful understanding of the selection criterion itself \citep{vitale_2021}. \cite{gerardi2021} corrected for GW anisotropy bias through modelling the dependence of successful GW detection on both BNS merger parameters and cosmological parameters, using general relativity (GR) and the GW detector's configuration. GR directly allows calculation of the strain and polarization breakdown of a GW produced by a BNS merger from that merger's measured parameters. Making a determination of detection on this GW further requires knowledge of the polarization sensitivity and strain detection threshold of the GW detector.

We extend this logic to EM anisotropy bias, using simulation-based inference (SBI) to characterize the EM selection criterion in a sample of BNS mergers. For an observed sample of BNS mergers detected in GWs, we expect to have both mergers with identified EM counterparts and those without. Mergers without EM counterparts in the sample allow us to infer the probability of EM detection for a BNS merger, given its measured parameters and assuming a model for the EM emission. This allows us to characterize the dependence of EM selection on BNS and cosmological parameters, and thereby correct for the EM anisotropy bias in standard siren measurements of $H_0$. The main innovation of our method is this inclusion of BNS mergers detected in GWs without detected EM counterparts.

In the absence of an analytic likelihood that includes both GW and EM selection effects, we intead opt for a SBI approach, where a BNS merger forward simulator enables emulation of both selection effects. SBI refers to a class of inference methods that rely on surrogate likelihood functions from a simulator forward model, enabling Bayesian inference even in situations where the likelihood function is intractable, such as ours \citep{Cranmer30055}. SBI methods typically function by simulating a data set from a set of input parameters, and then producing a summary statistic which describes the difference between the simulated data and the observed data. Through repeated sampling of the parameter space, these summary statistics are used to construct a surrogate likelihood function, and thus enable posterior inference. 

Here, we present an SBI-based method which corrects for both GW and EM anisotropy biases in standard siren measurements of $H_0$. Our approach exploits our singular goal of inferring the marginal posterior distribution of $H_0$, allowing us to treat \textit{all} BNS merger parameters as nuisance parameters. Specifically, we use marginal neural ratio estimation (MNRE), in which a summary statistic is generated for each set of input parameters. The summary statistic is a quantity which encapsulates the consistency of the parameters drawn with the observed data. These summary statistics are then used as a training set for a neural ratio estimator network, which learns the marginal posterior distribution for $H_0$, our parameter of interest  \citep{miller2021truncated}. 

MNRE is a recently-developed SBI method which requires far fewer samples than competing methods to produce an informative posterior. It accomplishes this by estimating only marginal posteriors rather than joint probability distributions \citep{miller2021truncated}. MNRE is thus appropriate in situations where only marginal distributions are of interest, and the computational expense of each sample is high. MNRE has recently been applied by \cite{sicret} in the context of Type Ia supernova cosmology, where it was used to marginalize over a large number of of supernova parameters. Similarly, our study is only interested in the marginal posterior distribution for $H_0$, and thus MNRE is an ideal tool.

To test and validate our SBI-based approach on a mock data set, we assume a set of true BNS mergers with associated GW strains. Of these BNS mergers, only a subset have associated EM detections. The expected kilonova light curves from the full merger sample is repeatedly simulated with randomly-sampled kilonova and cosmological parameter configurations, assuming a model for the kilonova emission, and the summary statistics from each round are used to train an MNRE. We demonstrate that by including GW-only mergers in a sample of BNS mergers, we can correct for the EM anisotropy bias latent within the sample. Our method provides a GW anisotropy bias correction level comparable to that in \cite{gerardi2021}, and further addresses EM anisotropy bias for standard siren cosmology.

The structure of this paper is as follows. Section \ref{sec:bias} explains the physical cause of each source of bias. Section \ref{sec:sim} discusses the mock data sets used in this study and the simulator. Section \ref{sec:method} provides a detailed discussion of our SBI-based method. Section \ref{sec:results} reviews results from each validation test performed on the mock data sets. We summarize and conclude in Section \ref{sec:conc}. 

\section{Sources of Bias}
\label{sec:bias}

\subsection{GW anisotropy and bias}
\label{subsec:gw}

\begin{figure}
    \centering
    \includegraphics[width=0.45\textwidth]{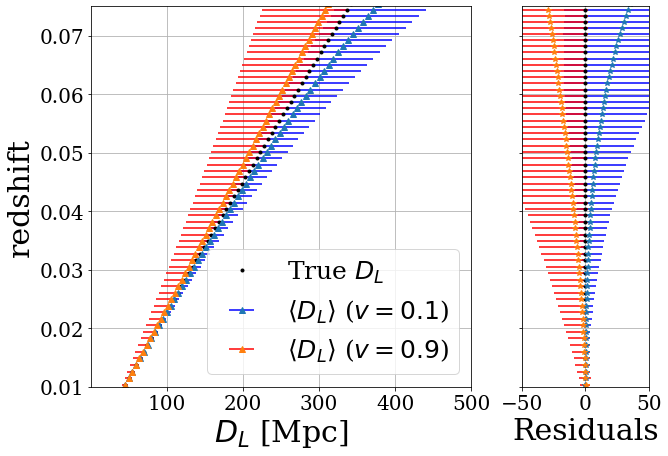"}
    \caption{Hubble diagram illustrating the source of gravitational wave GW anisotropy bias. Face-on mergers ($v=\cos(\iota)=0.9$) have inferred luminosity distances ($\langle D_L \rangle$) skewed nearer to the observer than their actual value, while the opposite is true for oblique mergers ($v=0.1$). The tendency to preferentially detect face-on mergers therefore results in an inflated value of the Hubble constant.}
    \label{fig:hchart}
\end{figure}

\begin{figure}
    \centering
    \includegraphics[width=0.45\textwidth]{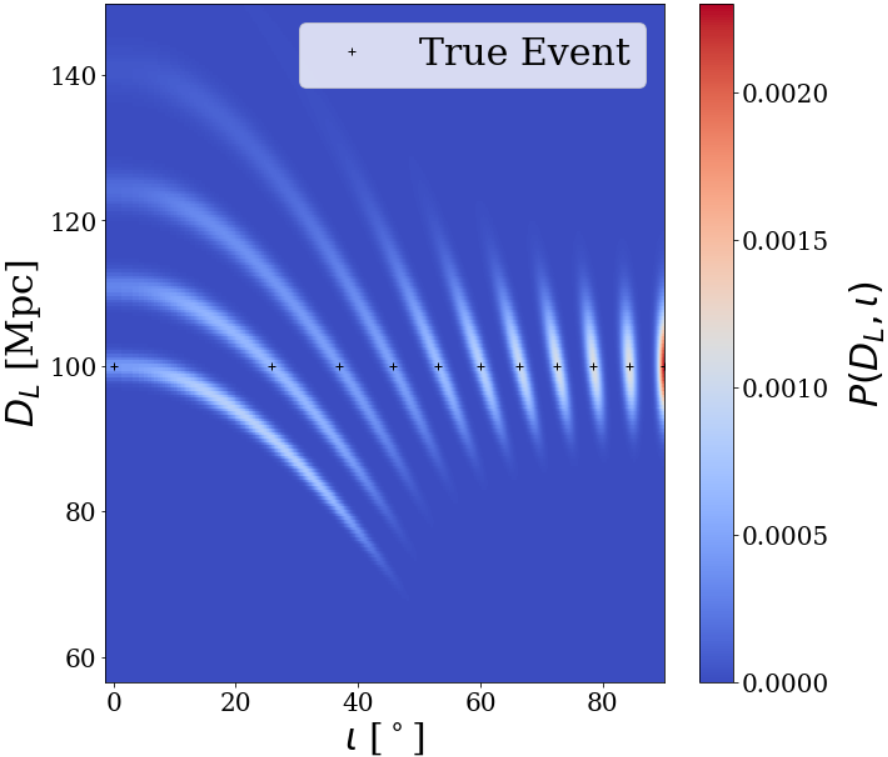}
    \caption{The joint $P(D_L,\iota)$ posterior distributions for ten simulated BNS mergers. Each merger is at exactly the same distance ($D_L=100$ Mpc) with the same NS masses (both $1.4$ M$_\odot$). Each merger varies only in its inclination angle, $\iota$. Mergers at low $\iota$ have highly degenerate posteriors, with significant probability density assigned to higher-than-true distances. These posteriors, when marginalized over $\iota$, are the source of the bias illustrated in Figure \ref{fig:hchart}.}
    \label{fig:pdl_example}
\end{figure}

GW detections of BNS mergers suffer from GW anisotropy bias, which results in standard siren measurements overestimating $H_0$ \citep{malmquist_bias}. This stems from the luminosity distance -- inclination angle degeneracy of the GW strain produced from a BNS merger, which skews the inferred luminosity distance of a BNS merger in a way that is dependent on its inclination angle $\iota$ \citep{gw_degeneracy}. We illustrate this effect in Figure \ref{fig:hchart}, where a series of BNS mergers evenly spaced in luminosity distance have their inferred luminosity distances and associated uncertainties shown for two possible inclination angles, one face-on and one oblique. Uncertainty in luminosity distance estimates from GW detections arises from the detector's polarization sensitivity \citep{cutler1994}. In Figure \ref{fig:hchart}, we assume the characteristics of the Advanced LIGO/Virgo experiment as expected in O4 \citep{yi2022gravitational}. Mergers placed at oblique inclinations have an inferred luminosity distance greater than their true value, while the opposite is true of mergers placed at face-on inclinations. This is because GW strain is strongest along the angular momentum axis of a BNS merger, and weakest perpendicular to this axis. A weak GW strain may be weak either because the merger is in fact very distant and face-on, or because the merger is nearby but at an oblique inclination, thus giving rise to the aforementioned degeneracy. An example of this luminosity distance-inclination angle degeneracy is shown in Figure \ref{fig:pdl_example}, which displays the overlaid $P(D_L,\iota)$ joint posterior distributions for $10$ otherwise identical BNS mergers at different inclination angles.

\begin{figure}
    \centering
    \includegraphics[width=0.45\textwidth]{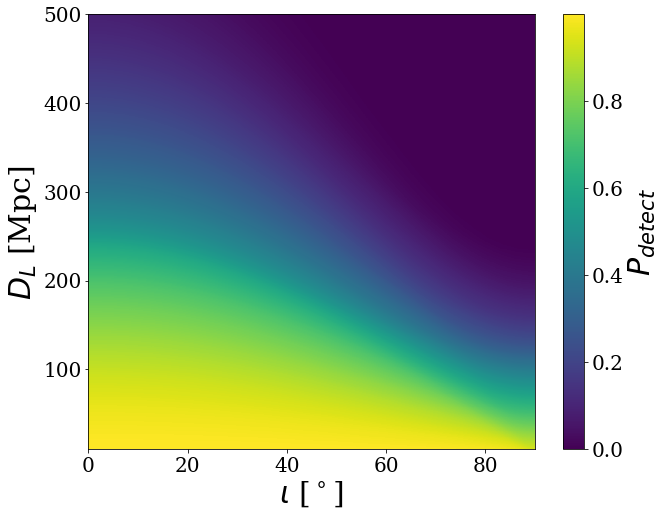}
    \caption{Probability of detecting GWs from a BNS merger given its luminosity distance, $D_L$, and inclination angle $\iota$. Probabilities were generated using \texttt{GWToolbox} assuming its parameterization of the Advanced LIGO/Virgo experiment expected in O4 \citep{yi2022gravitational}. Face-on mergers ($\iota\approx0^\circ$) are in general more likely to be detected than oblique mergers ($\iota\approx90^\circ$). This effect is especially significant for distant mergers.}
    \label{fig:gw_detect}
\end{figure}

This luminosity distance -- inclination angle degeneracy would be unproblematic in the absence of selection effects. However, the amplitude of a GW strain curve is related to the probability that the merger's GW is detected at all. In order for a BNS merger to be detected, its strain must exceed some critical signal-to-noise ratio. Naturally, this is less likely for weak-strain mergers than for strong-strain mergers. The nature of this relationship is shown in Figure \ref{fig:gw_detect}, where we show that face-on mergers are likely to be detected even at large distances, while oblique mergers are unlikely to be detected even at small distances. As a result, GWs from face-on mergers are preferentially detected in a sample of BNS mergers, biasing the luminosity distances to be nearer than their true values, and thus inflating the value of $H_0$ as inferred from standard sirens.

\begin{figure*}
    \centering
    \includegraphics[width= 0.8 \textwidth]{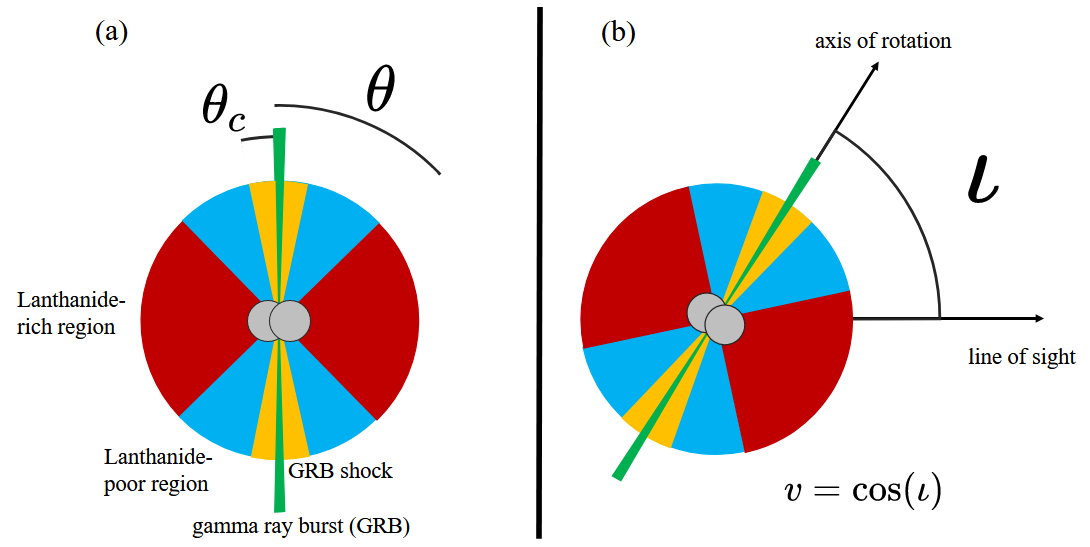}
    \caption{Geometry of a binary neutron star (BNS) merger and its associated kilonova. {\it Left panel} shows the geometry of the ejecta and emission, including the red and blue ejecta components, and the gamma-ray burst (GRB) shock cocoon. $\theta$ labels the half-opening angle of the blue component, and $\theta_c$ labels that of the GRB shock cocoon. {\it Right panel} demonstrates how the inclination angle, $\iota$, is defined with respect to the angualar momentum axis of the BNS. $v$, defined as the cosine of the inclination angle, is often used in the literature in lieu of $\iota$.}
    \label{fig:kn_geo}
\end{figure*}

\begin{figure*}
    \centering
    \includegraphics[width=0.95 \textwidth]{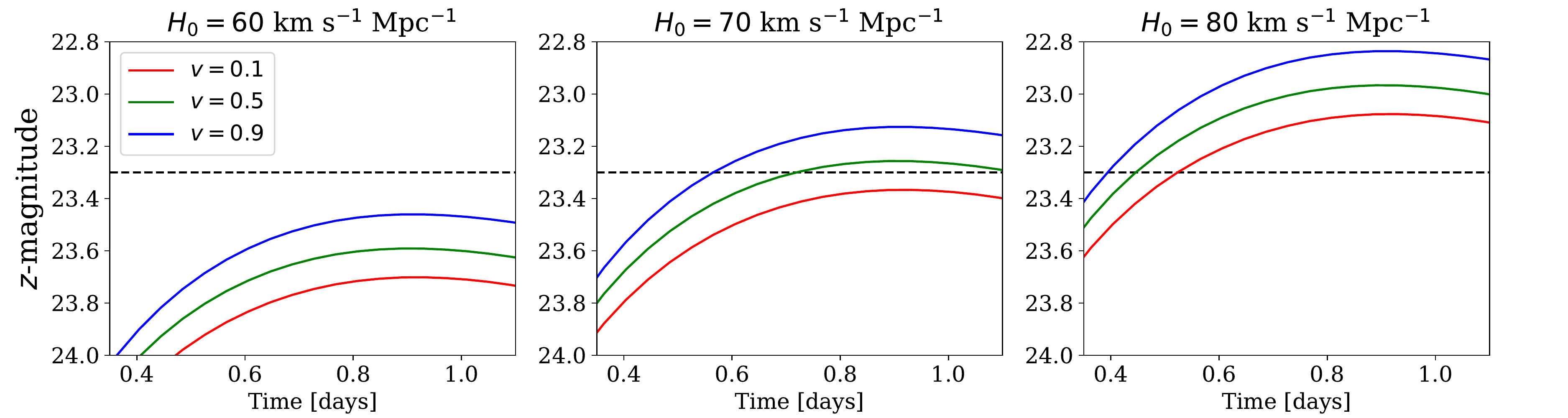}
    \caption{Simulated kilonova light curves, observed at various inclination angles, and in various cosmologies. Each light curve was generated using \texttt{MOSFiT} using the same parameter values, except for $H_0$ and inclination angle. When a kilonova is detected, it can be uncertain whether its detection was due to a high value of the $H_0$, or a low inclination angle.}
    \label{fig:lc_ex}
\end{figure*}

\subsection{EM anisotropy and bias}

A second source of bias affecting standard siren measurements arises from anisotropy of the kilonova EM emission and its associated selection effect. We provide a brief explanation for this bias below, and its expected effect on $H_0$. For a more in-depth discussion of the origin and nature of this bias, we refer to \cite{chen2020}. 

BNS mergers produce an associated kilonova, whose EM emission enables multi-messenger standard siren cosmology. This emission primarily arises from $r$-process nucleosynthesis in the neutron-rich ejecta of the BNS merger. Simple kilonova models often invoke two ejecta components, a `blue' polar component, and a `red' equatorial component. The geometry of the emission from these components is schematically depicted in Figure \ref{fig:kn_geo}. Following \cite{nicholl2021} and \cite{bulla2019}, we characterize this geometry using the blue component half-opening angle, $\theta$. The exact value of this angle is uncertain, with \cite{bulla2019} estimating $\theta=30^\circ$ for GW170817 and \cite{nicholl2021} suggesting that $\theta=45^\circ$ is an appropriate guess for all kilonovae.

As a result of the anisotropic EM emission from the kilonova, the inclination angle of a BNS merger influences whether or not the kilonova can be detected in EM telescope follow-up observations. When a BNS merger is detected in GWs but its kilonova is not discovered, the observer does not know whether the kilonova was missed due to excessive distance or inclination. This issue is illustrated in Figure \ref{fig:lc_ex}, where kilonova light curves from a merger at a fixed redshift are produced for various assumed $H_0$ and inclination angles. This figure demonstrates that identical BNS mergers can fail to produce a detectable EM signal due to either the assumed value of $H_0$, or its inclination. As a result, this effect can cause the observer to preferentially discover face-on kilonovae in optical imaging follow-up searches. Since this EM anisotropy bias further enforces the discovery of standard sirens whose inferred luminosity distances are nearer than their true values, this again inflates the inferred value of $H_0$. Thus, EM selection acts to exacerbate the already extant GW anisotropy bias in a sample of BNS mergers.

\section{Mock Data Sets}
\label{sec:sim}

To develop and validate our approach, we use a mock GW data set of sample of BNS mergers, as well as accompanying simulated EM light curves for a subset of these mergers for which the kilonova is detected. A GW detection provides three relevant BNS parameter distributions: (1) the joint luminosity distance -- inclination angle distribution $P(D_L,\iota)$, (2) the joint NS mass distribution $P(M_1,M_2)$, and (3) the GW sky localization $P(\text{RA},\text{DEC})$. An EM detection of the kilonova counterpart provides a redshift $z$, and precise sky location $(\text{RA}, \text{DEC})$.

For each merger, the joint probability distribution for the inclination angle and luminosity distance is the related to the merger's true values by the relation

\begin{multline}
    dp(v,D_L)=\mathcal{N}\left(\frac{D_L}{D_0}\right)^2 \\ \times\text{exp}\left[-\frac{1}{2\Delta_1^2}\left(\frac{v D_0}{D_L}-v_0\right)^2-\frac{1}{2\Delta_2^2}\left(\frac{D_0(1+v^2)}{2D_L}-\frac{1+v_0^2}{2}\right)^2\right]\\ \times \Theta(D_L/D_0)\Theta\left(\frac{D_\text{max}}{D_0}-\frac{D_L}{D_0}\right)\Theta(1-v^2)dvdD_L,
\end{multline}

\noindent where $v=\cos(\iota)$ is the inferred cosine of the inclination angle, $D_L$ is the inferred luminosity distance, $D_0$ is the true luminosity distance, $v_0=\cos(\iota_0)$ is the true cosine of the inclination angle, $\Delta_1$ and $\Delta_2$ are functions which encode the polarization sensitivity functions of the GW detector, $\Theta$ is the Heaviside step function, and $D_\text{max}$ is an arbitrarily high distance which is treated as the cutoff for the probability distribution \citep{cutler1994}. In this study, we use a fiducial $D_\text{max}=6.5$ Gpc. While the inclination angle is often marginalized over to convert this into a luminosity distance posterior distribution, it is important for us to leave them separate, as our goal is to discern the bias produced by the inclination angle.

We perform two kinds of validation tests, each depending on a different underlying mock sample of BNS mergers. The first class of tests are \textit{contrived tests}, where the parameters of each BNS merger are conspicuously chosen to highlight a certain effect. For example, a sample of $100$ mergers may be placed at the same true distance and redshift, but each with different inclination angles, to exaggerate and test the effect of EM inclination angle selection bias. The second class of tests are \textit{cosmological tests}, where the BNS parameters are simulated using \texttt{GWToolbox}, a toolkit for generating realistic GW source populations and their probability of detection with various GW instruments \citep{yi2022gravitational}. In our cosmological tests, we treat all mergers as having been detected by LIGO/Virgo during O4.

\texttt{GWToolbox} generates a set of true parameters for each merger, produces a strain curve from those parameters, and then verifies that the strain's signal-to-noise ratio (SNR) is high enough for the merger to be detected. If detected, the strain curve is then interpreted to produce BNS parameter estimates. In this study, we set the detection SNR to $8$. Mergers generated using \texttt{GWToolbox} will thus suffer from GW anisotropy bias due to the dependence of the probability of detection on luminosity distance and inclination angle, as illustrated in Figure \ref{fig:gw_detect}. This stands in contrast with the contrived mergers, where a sample of mergers is assumed to be detected regardless of their parameters, and is therefore not subject to GW anisotropy bias. Contrived mergers are therefore only useful in tests where EM-selection bias is considered independently of GW anisotropy bias.

Once a sample of GW mergers is produced through either method, the telescope follow-up imaging must be specified for each merger. In our simplest case, it is assumed that a telescope is always available for EM follow-up and it is always pointed in the right location in the sky to detect the kilonova. We also assume that the exposure depths of the images are the same for each merger. In this case, EM follow-up can only fail if the kilonova is too dim to be detected in the exposure depths of the images. A merger may be too dim either because it is too distant, or because its inclination angle is too oblique, thus giving rise to EM-selection bias. This method is employed for the contrived tests in this study.

In the more complex but realistic case, the precise sky localization of the BNS merger is considered. Given the true location of a BNS merger on the sky and the telescope imaging pointings, there exists some probability that the merger lies outside all pointings and could not have been observed. Furthermore, even if the merger indeed lies within a pointing, the depth, time post-merger, and filter of that image, as well as the Galactic dust extinction at that sky location, must all be considered.  This method is employed for the cosmological tests in this study.

In our validation tests including sky localization, we assume the same GW sky localization and EM follow-up for each merger. One test uses the GOTO-4 follow-up for GW190425 \citep{Steeghs_2022}, which is a very incomplete follow-up tiling (having only $29\%$ probability coverage of the GW localization region), while another uses the CFHT follow-up for GW 190814 to contrast with a deeper and more complete tiling with $64\%$ probability coverage \citep{Abbott_2020,Abbott_2020b,Vieira_2020}.

The Galactic dust extinction at a BNS merger's sky location will also affect the probability of EM detection of the kilonova, as a kilonova in a sky region with high dust extinction is less likely to be detected than one in a region with low dust extinction. We use the Galactic dust maps of \citet{dust_recalibration} through the \texttt{dustmaps} package. In contrived tests, wherein sky plane sampling is not considered, dust extinction is set to a fiducial value of $E_{B-V}=2.2$.

When evaluating whether or not a kilonova should be detected, its light curve must be generated. We generate kilonova light curves for each simulated merger from its BNS parameters using \texttt{MOSFiT}, a software package for astrophysical transient simulation \citep{Guillochon_2018,nicholl2021}. If the kilonova's magnitude is brighter than the exposure depth of the relevant telescope pointing, then the EM counterpart is detected. In this way, each BNS merger in the merger sample is evaluated and classified either as an `EM+GW' merger (EM counterpart detected) or a `GW-only' merger (EM counterpart not detected). Leveraging information from both EM+GW mergers and GW-only mergers to correct for EM selection bias is a core feature of this study.

\section{Method}
\label{sec:method}

\begin{figure*}
    \centering
    \includegraphics[width= 0.7 \textwidth]{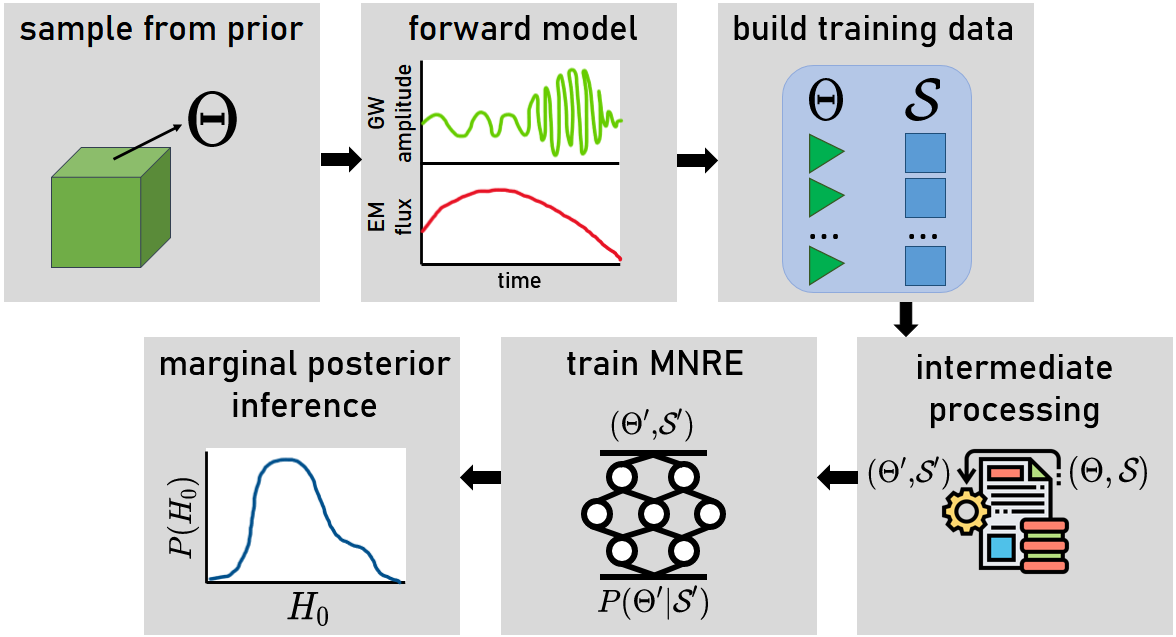}
    \caption{Overview of our approach to inference. EM and GW data supplied to the simulator informs the parameter prior distributions set by the model. In practice, this should be real data, but in this study we use a simulated mock data set. At each step of simulation, BNS and cosmological parameters are sampled from these prior distributions and passed to a forward model. The forward model simulates the BNS mergers assuming these parameters and determines whether their GW and EM emission should be detected. This is done repeatedly to build a set of training data. The data then undergoes intermediate processing to allow for the correction of EM anisotropy before being passed into the MNRE for training. A fully-trained MNRE outputs a marginal posterior distribution for $H_0$.}
    \label{fig:method_flowchart}
\end{figure*}

\begin{figure*}
    \centering
    \includegraphics[width= 0.7 \textwidth]{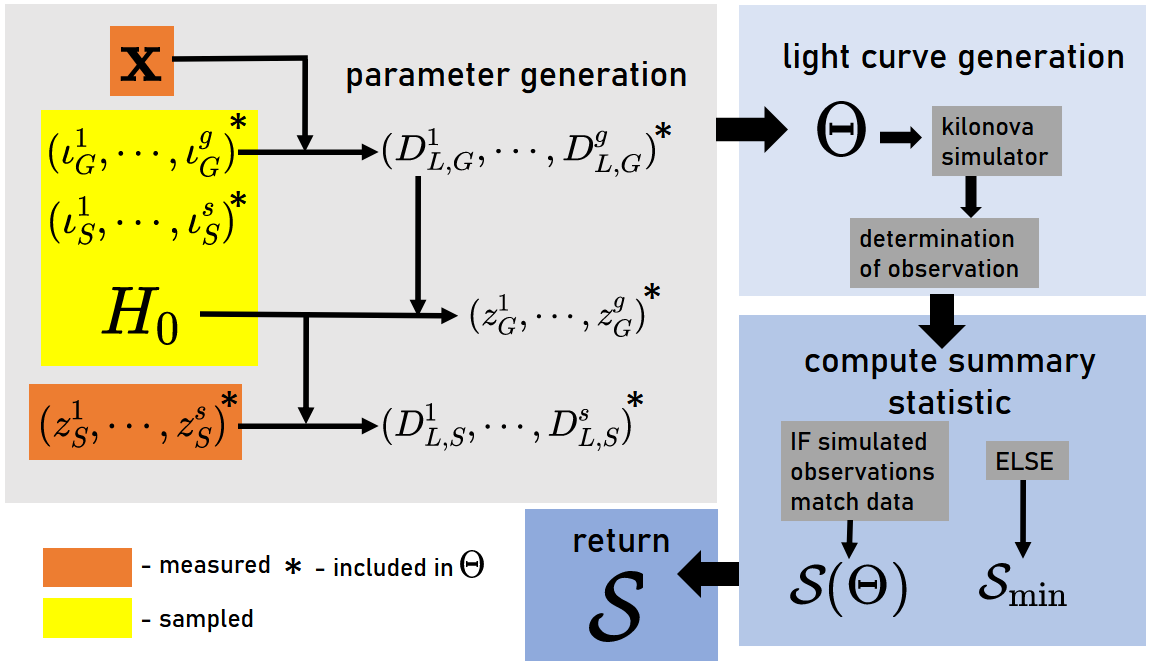}
    \caption{Process overview of the forward model developed for this study. In the first step, the parameters necessary for BNS merger generation are sampled and computed. The inclination angles, $\iota_G$ and $\iota_S$, as well as the Hubble constant, $H_0$, are sampled from a prior distribution. $\mathbf{x}$ represents the GW strain data, which is used to generate a random luminosity distance for GW-only mergers, $D_{L,G}$. Meanwhile, the luminosity distances of standard siren mergers, $D_{L,S}$, is constrained by the sampled $H_0$ and the measured redshifts, $z_S$. The redshifts of GW-only mergers are similarly computed from $D_{L,G}$ and $H_0$. These parameters are summarized as $\Theta$, which is passed to \texttt{MOSFiT} for light curve generation. Each light curve is either observed or not observed according to some magnitude criterion. If all EM+GW mergers are confirmed to have been EM-observed and all GW-only mergers are not EM-observed, then the simulated observations are said match the data, and a summary statistic is computed. Otherwise, the minimum, or `null', summary statistic is returned.}
    \label{fig:forward_flowchart}
\end{figure*}

\subsection{The Case for SBI}
\label{sec:case}

In this section we discuss the insufficiency of traditional inference methods in accounting for EM and GW anisotropy biases. We begin by laying out the traditional approach to inferring parameters from sets of BNS mergers. To simplify this discussion, we neglect the inference of BNS parameters to focus solely on $H_0$. Given a fixed catalogue of mergers, $\mathbf{x}$, with estimates on $D_L$, $z$, and $\iota$, we may write the posterior distribution for $H_0$ as

\begin{multline}
    P(z,D_L,\iota,H_0|\mathbf{x})\propto\frac{P(H_0)}{[\bar{N}(H_0)]^N}\times \\ \prod^N_{i=1}P(\hat{z}_i|z_i,H_0,D_L)P(\hat{D}_{L,i},\hat{\iota}_i|D_{L,i},\iota_i),
    \label{eq:anpost}
\end{multline}

\noindent where $\hat{z}_i$, $\hat{D}_{L,i}$, and $\hat{\iota}_i$ are the estimates of $z_i$, $D_{L,i}$, and $\iota_i$ for each merger, $\bar{N}$ is the mean number of detected EM+GW mergers as a function of $H_0$, and $N$ is the number of EM+GW mergers in the catalogue \citep{Mortlock_2019}. No analytic formula exists to produce $\bar{N}$ as a function of $H_0$; for any given catalogue of mergers, each merger's luminosity distance and inclination angle influence both the probability of GW detection and the probability of EM detection, as discussed in Section \ref{sec:bias}.

A brute force estimation of $\bar{N}$ for a given $H_0$ requires generating the total number of mergers within a volume of space over some duration of time, and then determining which among them would be detected as a standard siren. The determination of detection requires simulating each merger in the catalogue and then comparing the expected GW and EM emission to both the response functions of the GW detector and the telescope tiling of the localization region. Since each merger has a unique luminosity distance and inclination angle, the number of parameters influencing the number of detected mergers in a given sample, $N$, scales with the number of mergers included within that sample. Estimating $\bar{N}$ at a particular $H_0$ furthermore requires a dense sampling of this parameter space, and an estimation of $\bar{N}(H_0)$ requires dense sampling assuming various values of $H_0$. Computational expenses mount as the required number of samples increases, rendering this approach prohibitively computationally expensive beyond a merger sample of more than a few mergers.

We therefore adopt an SBI approach to estimating $\bar{N}(H_0)$. Within the SBI framework, we repeatedly sample the parameter space and simulate a sample of BNS mergers at each point. The mock observables from these `forward simulations' enable inference of the parameter of interest, in this study, $H_0$. We train a neural network to compare the mock observables to real data and thereby learn the parameter values underlying that data. Our method employs marginal posterior inference, completely bypassing calculation of the likelihood function and thus removing the need for explicit knowledge of $\bar{N}(H_0)$ \citep{malmquist_theory}.

\subsection{Layout of Approach}

Our SBI method follows a six-step approach:

\begin{enumerate}
    \item Draw parameter sample $\Theta$
    \item Generate realistic observables, $\mathbf{x}|\mathbf{x}_0,\Theta$
    \item Summarize consistency of generated observables with data using a summary statistic $\mathcal{S}=f(\mathbf{x},\mathbf{x}_0)$
    \item Repeat steps i-iii to produce data set $(\Theta,\mathcal{S})$
    \item Intermediate processing to prepare data set for training
    \item Train neural network to infer the marginal posterior of $H_0$ from $(\Theta,\mathcal{S})$.
\end{enumerate}

\begin{table}
    \centering
    \begin{tabular}{|l|l|l|}
    \hline
    \textbf{Parameter} & \textbf{Prior} & \textbf{Motivation}       \\ \hline
    $H_0$     & Uniform{[}60, 80{]}                       & Treat as unconstrained  \\ \hline
    $D_L$     & $P(D_L,\iota|\mathbf{x}_{GW})$            & Inferred from GW strain \\ \hline
    $\iota$   & $P(D_L,\iota|\mathbf{x}_{\text{GW}})$     & Inferred from GW strain \\ \hline
    (RA, DEC) & $P(\text{RA},\text{DEC}|\mathbf{x}_{GW})$ & From GW LAL Inference \\ \hline
    \end{tabular}
    \caption{The four free parameters considered in this study and their prior distributions. A GW detection providing strain $\textbf{x}_{\text{GW}}$ produces a joint posterior distribution for the luminosity distance and inclination angle, $P(D_L,\iota|\textbf{x}_{\text{GW}})$. This distribution is treated as the credible region for an merger's $D_L$ and $\iota$. Similarly, a GW detection has an associated sky localization, $P(\text{RA},\text{DEC}|\textbf{x}_{\text{GW}})$, produced by LAL inference of the GW strain.}
    \label{tab:prior-ref}
\end{table}

\noindent Below, we discuss each step in greater detail, proceeding in order from i to vi. Figure \ref{fig:method_flowchart} displays a schematic of our approach.

In step (i), the forward model gathers a sample of parameters from their relevant prior space. These parameters include the luminosity distances and inclination angles of each BNS merger, as well as $H_0$. During tests considering the impact of telescope follow-up we also include the sky location, $(\text{RA},\text{DEC})$. Table \ref{tab:prior-ref} summarizes the appropriate minimally-informative prior distributions.

After gathering the parameter sample $\Theta$, the forward model generates the GW and EM observables for each event (step (ii)). Since training a neural network directly on kilonova light curves and merger GWs is difficult, we introduced a data compression scheme to aid training convergence (for a similar example in cosmology, see \citealt{Alsing_2018}). We perform this compression in steps (iii) and (v). Step (iii) computes the similarity between the simulated and actual observables, producing a quantity called the summary statistic $\mathcal{S}$.

In step (iv), the forward model repeatedly samples parameter space and generates observables, thus producing a data set of input parameters, $\Theta$, and their associated summary statistics, $\mathcal{S}$. This data set, $(\Theta,\mathcal{S})$, contains information on the posterior distributions of the input parameters, including $H_0$. In this work, we train a neural network to reconstruct the marginal posterior distribution $P(H_0)$ using the data set produced by the forward model. However, we found that the noise inherent to our choice of summary statistic hindered training convergence. We therefore apply intermediate processing in step (v), wherein we recast the data to a basis more amenable to training convergence. This transformation varies with the global quantities of the dataset (e.g., maximum and minimum), so we apply it after the completion of the sampling and forward modelling phase. Then, in step (vi), a marginal neural ratio estimator (MNRE) trained on these transformed data produces the marginal posterior distribution for $H_0$.

\subsection{Forward Model}
\label{sec:forward}

\subsubsection{Observable Generation}

The forward model generates observable data, $\mathbf{x}$, given a parameter vector, $\Theta$. Our input parameters consist of $H_0$, an inclination angle $\iota$ for each merger, and a luminosity distance $D_L$ for each GW-only merger. The observables in this study consist of two parts: GW emission and EM emission.

The GW strain for a BNS merger provides a joint estimate for its $D_L$ and $\iota$. Example joint $(D_L,\iota)$ distributions are shown in Figure \ref{fig:pdl_example}. For an EM+GW merger, the measured redshift $z$ and sampled $H_0$ specify a $D_L$. We then sample an inclination angle $\iota$ from the corresponding conditional posterior distribution, $P(\iota|D_L,x_{\text{GW}})$, where $x_{\text{GW}}$ represents the GW strain for that merger. Meanwhile, a GW-only merger lacks a measured redshift, leaving both $D_L$ and $\iota$ as free parameters to sample from the joint posterior distribution $P(D_L,\iota|x_{\text{GW}})$. Following these rules, the forward model fixes the parameters of all mergers in the sample prior to light curve generation. For a graphical representation, see the \texttt{parameter generation} panel of Figure \ref{fig:forward_flowchart}.

Before generating light curves, we produce GW observables for each merger. The forward model uses \texttt{GWToolbox} to produce a GW strain and detection signal-to-noise ratio for each merger in the sample. We count the GW emission of a BNS merger as detected if the merger's SNR exceeds $8$. Only GW-detected events contribute to parameter inference. Such events supply the GW-inferred joint posterior distribution for $D_L$ and $\iota$.

The forward model then generates realistic EM emission (kilonovae) for each GW-detected merger using the \texttt{MOSFiT} software package. \texttt{MOSFiT} uses a number of parameters to specify the features of a kilonova, for a detailed description of these parameters see \cite{nicholl2021}. While we sample $D_L$ and $\iota$ from their prior distributions and pass those samples to the light curve simulator, we fix all other relevant parameters to fiducial values. Appendix \ref{apdx:params} discusses our assumed values for the following kilonova parameters: the kilonova ejecta component opacities, $\kappa_\text{red}$ and $\kappa_\text{blue}$, the neutron star masses, the disk ejection fraction $\epsilon_\text{disk}$, the blue ejecta enhancement factor $\alpha$, and the geometric parameters $\theta$ and $\theta_c$, which respectively describe the half-opening angles of the blue ejecta component and gamma ray burst-shocked region.

\subsubsection{Summary Statistic}

The summary statistic $\mathcal{S}$ represents the similarity of the simulated observables with the observed data, $\mathbf{x}$. Since the simulated observables are generated from the parameter sample $\Theta$, the summary statistic captures the \textit{consistency} of $\Theta$ with $\mathbf{x}$.

For samples where the simulated and observed data do not meet a minimum consistency threshold, we set the summary statistic to a minimum or `null' value. We refer to such samples as `non-informative' since they do not contribute positively to the inferred posterior density. Our minimum consistency check ensures that only the correct mergers within a sample have detected EM counterparts. 

To illustrate this `minimum consistency check' with a basic example, consider a real set of $2$ BNS mergers where $1$ has a detected EM counterpart. Following our procedure, we sample a parameter sample from prior space and produce mock GW and EM observables from that sample. Should zero or both BNS mergers have detected EM counterparts, then we say that the simulated observables are completely inconsistent with the real data, and a null return is produced. Furthermore, if the wrong merger has a detected EM counterpart in the simulated data, then that is also completely inconsistent, producing a null return.

For samples where the minimum consistency threshold is met, we then assess the \textit{degree} of consistency. We do this by comparing the sampled $D_L$ and $\iota$ to the $(D_L,\iota)$ joint posterior distribution inferred from each merger's GW signal. Appendix \ref{apdx:sumstat} discusses the exact form of this comparison, as well as our choice of minimum return for non-informative samples.

\subsubsection{EM Follow-Up}

We investigate how imperfect EM follow-up observation attempts influence our ability to correct for EM anisotropy bias. To motivate this, consider a scenario where we can image the full sky to some exposure depth in some filter for several days after a BNS merger is detected in GWs. In this scenario, the only cause for non-detection of the kilonova is the flux from the kilonova does not meet the detection thresholds of the telescope images. Insufficient flux can only be explained by either the distance of the merger or its inclination, both features of the underlying BNS merger. Therefore, the EM non-detection places some constraints on these BNS merger parameters. The existence of these constraints allows for the correction of EM anisotropy bias.

Let us now consider the case where a realistic telescope follow-up is performed. When a BNS merger is detected in GWs, LAL inference is used to localize its origin on the sky with a probability density map \citep{Veitch_2015}. Telescope follow-up tiling of the localization region is not exhaustive, as they do not cover $100\%$ of the sky, nor are the exposures always deep enough to guarantee kilonova detection. This introduces new failure modes for EM detection: the possibility that an merger was not within the field of view of any images during the follow-up campaign, or the images were not of sufficient depth. For any given EM non-detection, it is therefore unclear whether the non-detected was due to factors intrinsic to the BNS merger ($D_L$, $\iota$) or factors relating to the telescope follow-up.

\subsection{Intermediate Processing}
\label{sec:emc}

We refer to the process of repeatedly simulating observables from parameter samples as a `run' of the forward model. A run of the forward model produces a table of input parameter vectors, $\Theta$, and their associated summary statistics, $\mathcal{S}$. These tables contain the information necessary to produce a marginal posterior distribution for $H_0$.

To correct for EM anisotropy bias, we use information from GW-only mergers to estimate the bias implicit in EM+GW mergers. We do this by inspecting the shift in summary statistics when we include all mergers in the sample versus when the sample only includes EM+GW mergers. Our approach to EM anisotropy correction requires two full runs of the forward model. We refer to the process whereby we graft the output of one run onto the other as `intermediate processing'.

In the first run, the data set includes only EM+GW mergers, and the forward model generates a sample of EM+GW summary statistics called $\mathcal{S}_{\textrm{S}}$. In the second run, the data set includes GW-only mergers along with EM+GW mergers, and the forward model generates a second set of summary statistics called $\mathcal{S}_{\textrm{GW}}$. The null statistics in $\mathcal{S}_\textrm{S}$ indicate $\Theta$ for which one or more mergers lack a detected EM counterpart. Meanwhile, the null statistics in $\mathcal{S}_\textrm{GW}$ may also indicate $\Theta$ for which one or more mergers erroneously have a detected EM counterpart. This additional constraint marginalizes over the EM selection effect, debiasing the $H_0$ posterior prescribed by the summary statistics. However, the uncertainties in the GW-only mergers' redshifts produce greater scatter in $\mathcal{S}_\text{GW}$ than in $\mathcal{S}_S$, effectively prescribing a much broader posterior distribution for $H_0$. We therefore process both sets of summary statistics into a modified summary statistic set $\mathcal{S}'$, which possesses the unbiasedness of $\mathcal{S}_\text{GW}$ and a precision nearing that of $\mathcal{S}_\text{S}$. We discuss the details of this `intermediate processing' step in Appendix \ref{apdx:ip}.

\subsection{Inference Model}
\label{sec:inference}

The inference model learns the marginal posterior distribution of $H_0$ from the set of input parameter vectors and their associated summary statistics produced by the forward model. We perform this inference using Marginal Neural Ratio Estimation (MNRE), a simulation-based inference approach \citep{miller2021truncated}. MNRE learns the marginal likelihood-to-evidence ratio for a parameter of interest from a set of training data consisting of input parameters and their associated model outputs. Appendix \ref{apdx:nre} describes in more detail neural ratio estimation, the operating principle behind MNRE.

\begin{table}
    \centering
    \begin{tabular}{|l|l|l|l|}
    \hline
    \textbf{Parameter} & \textbf{EM-Only} & \textbf{GW-Only} & \textbf{Combined} \\ \hline
    $l_i$ & $5\cdot10^{-4}$ & $5\cdot10^{-5}$ & $5\cdot10^{-4}$  \\ \hline
    $f$ & $0.1$ & $0.1$ & $0.5$ \\ \hline
    $p$ &  $5$ & $10$ & $20$ \\\hline
    $e_m$ & $25$ & $50$ & $100$ \\ \hline
    \end{tabular}
    \caption{The training hyperparameters used in neural ratio estimation (NRE) for each test. These are the initial learning rate $l_i$, the learning rate adjustment factor, $f$, the patience, $p$, and the maximum number of epochs $e_m$. The NRE begins training at the initial learning rate, which is then adjusted by the learning rate scheduler as training progresses. At each step in training, a validation loss is calculated and passed to the learning rate scheduler. If $p$ rounds pass without any decrease in validation loss, then the learning rate is multiplied by the factor $f$. This continues for $e_m$ epochs.}
    \label{tab:hyperpar}
\end{table}

We train the neural network using \texttt{Adam} \citep{adam} to minimize binary cross-entropy loss, as defined in \citet{miller2021truncated}. Although MNRE tends to produce conservative posterior estimates once it converges, inappropriate selection of training hyperparameters often leads to convergence issues \citep[e.g.,][]{lratio1, Brehmer_2018, Brehmer_2019}. To produce reliable posteriors in each training, we adjusted the learning rate schedule and other hyperparameters, as shown in Table \ref{tab:hyperpar}. We implement MNRE using the \texttt{swyft} software package \citep{miller2020simulation,miller2022}.

\section{Results on Mock Merger Data Sets}
\label{sec:results}

\begin{figure*}
    \centering
    \includegraphics[width=\textwidth]{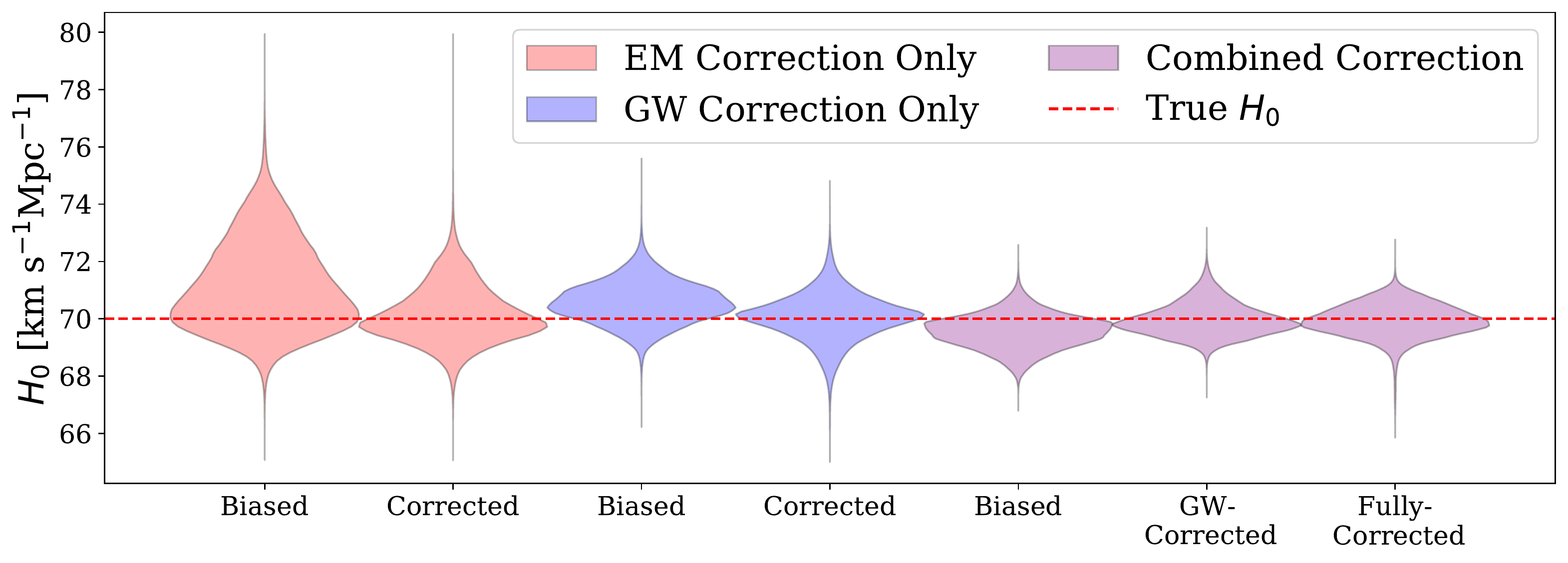}
    \caption{The $H_0$ posteriors produced by each of the first three tests in this paper. The posteriors are coloured according to their test, and labelled according to their status as biased, corrected, or partially-corrected. In the EM-correction only and GW-correction only tests, two posteriors are produced, one biased and one corrected. In the combined tests, one with and one without sky-sampling considerations, three posteriors are produced, one biased, one with GW anisotropy bias corrected, and one with all biases corrected. It is meaningless to correct for EM anisotropy bias without correcting for GW anisotropy bias if the data suffers from GW anisotropy bias, so those posteriors were not produced in this study. The combined test results pictured here correspond to the C1 test in Table \ref{tab:combined_tests}.}
    \label{fig:violin_results}
\end{figure*}

\begin{figure*}
    \centering
    \includegraphics[width=\textwidth]{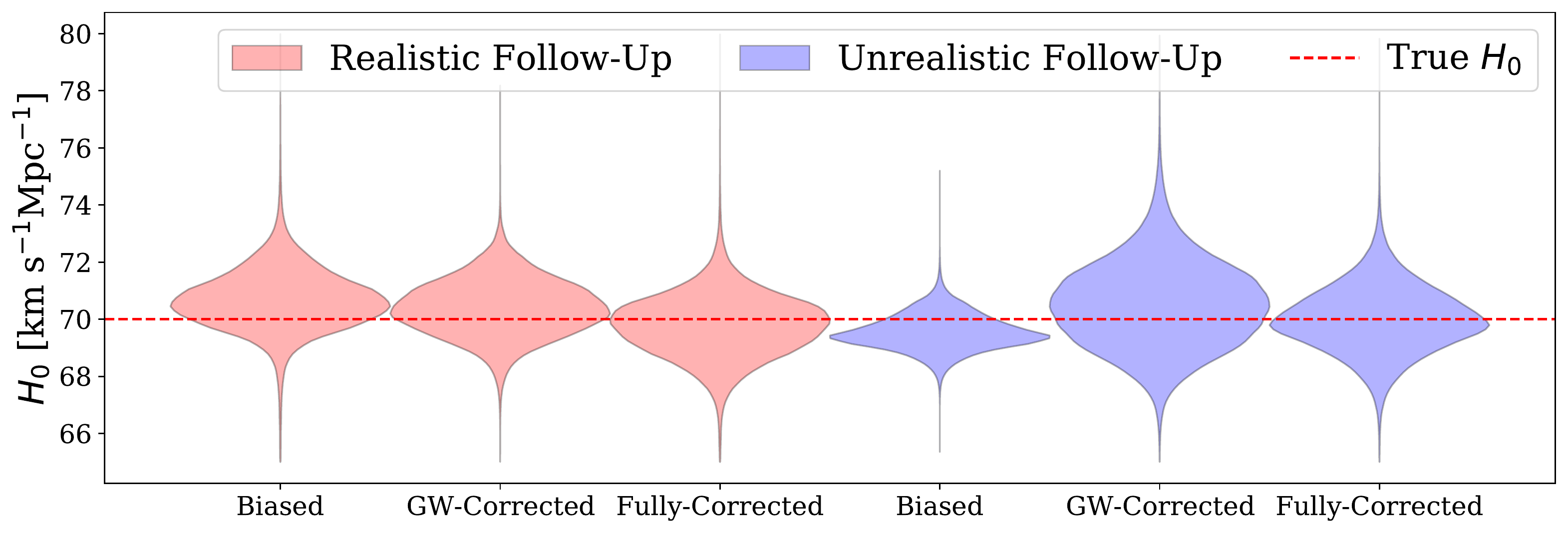}
    \caption{The $H_0$ posteriors produced by increasing the EM detection sensitivity on the combined EM+GW correction test. The posteriors are coloured according to their test, and labelled according to their status as biased, corrected, or partially-corrected. Even when realistic follow-up is considered, EM and GW anisotropy bias can be corrected for using our method, given that the EM detection sensitivity is sufficiently high. With reference to Table \ref{tab:combined_tests}, the unrealistic follow-up test corresponds to C3 and the realistic follow-up test corresponds to C5.}
    \label{fig:violin_results2}
\end{figure*}

\subsection{GW-only Correction}

\begin{figure}
    \centering
    \includegraphics[width=0.45\textwidth]{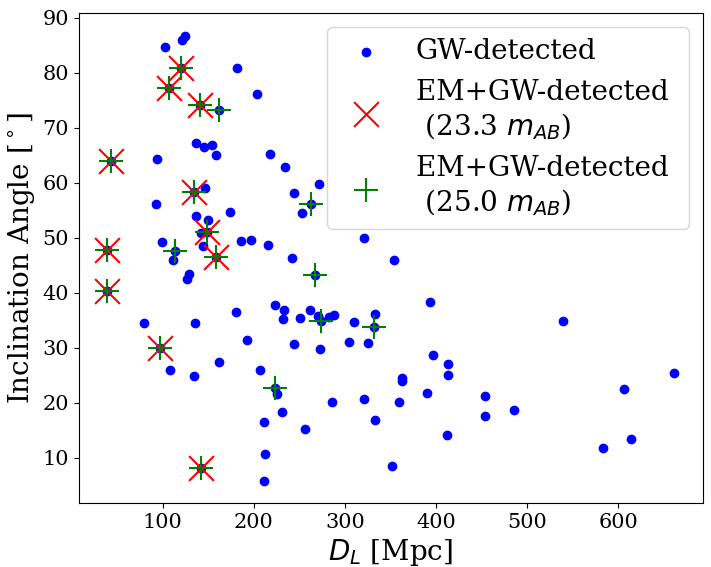}
    \caption{The luminosity distances ($D_L$) and inclination angles of the 100 GW-detected BNS mergers generated using \texttt{GWToolbox}. Mergers at large distances are unlikely to be detected unless they are nearly face-on (see Figure \ref{fig:gw_detect}), thus resulting in GW anisotropy bias. Mergers which would be detected via EM follow-up with a $g$-band magnitude cutoff of $23.3$ are marked in red, and mergers which would be detected with a cutoff of $25.0$ are marked in green.} 
    \label{fig:100_dl}
\end{figure}

In this test, 100 simulated BNS mergers were generated using \texttt{GWToolbox} assuming the realistic `cosmological' parameter distribution discussed in Section \ref{sec:sim}. This sample of mergers suffers from GW anisotropy bias, as distant mergers are only detected if their inclination angles are nearly face-on. This is illustrated in Figure \ref{fig:violin_results}. For this test, the forward model was modified to not evaluate light curves at all, and to always assume that a GW-detected merger is also an EM+GW merger regardless of its luminosity distance or inclination angle. This is equivalent to assuming that all mergers lie within the field of view of a telescope with infinite exposure depth. The purpose of this modification is to ensure that GW anisotropy bias is corrected in the absence of EM anisotropy bias. In essence, this test is equivalent to the GW anisotropy bias correction test discussed in \cite{gerardi2021}.

Two sets of summary statistics are generated using this forward model. In one set, called $\mathcal{S}_\text{b}$, GW anisotropy is not considered, and the full sample of 100 mergers are considered in each summary statistic. In the other set, called $\mathcal{S}_\text{c}$, GW anisotropy is considered, so only a subset of the 100 mergers are considered in each summary statistic, and this subset varies in size and composition from one draw to the next. The first of these sets is used to produce a biased $H_0$ posterior, while the second produces the corrected $H_0$ posterior.

These posteriors are shown in Figure \ref{fig:violin_results}. The value of $H_0$ predicted by the biased distribution is $70.54^{+0.73}_{-0.69}$ km s$^{-1}$ Mpc$^{-1}$, which indicates a GW anisotropy bias of $0.54^{+0.73}_{-0.69}$ km s$^{-1}$ Mpc$^{-1}$. The corrected distribution yeilds $H_0=70.09^{+0.69}_{-0.76}$ km s$^{-1}$ Mpc$^{-1}$, which accounts for $83.33\%$ of the bias. The bias in our distribution is within two standard deviations of the mean bias computed by \cite{gerardi2021}, indicating that our method for merger generation and bias measurement is consistent with their results. The degree to which our method corrects for the bias in $H_0$ is also comparable to their results, which produce corrected distributions with biases consistent with $0$ to the $1$-$\sigma$ level.

\subsection{EM-only Correction}
\label{sec:10ce}




In this test, 10 BNS mergers occur at the same luminosity distance and redshift. This distance is set to $135.9$ Mpc, which has a corresponding redshift of $0.031$ assuming $H_0=70$ km s$^{-1}$ Mpc$^{-1}$. Each merger is identical except for its inclination angle, $\iota$. Every neutron star is assumed to have a mass of $1.4 M_\odot$. The inclination angles of the mergers are evenly spaced in $\cos(\iota)$, taking values from $\cos(\iota)=0.0$ to $\cos(\iota)=0.9$. Of these, only the three mergers with the lowest $\iota$ are EM+GW mergers. By placing these mergers at the same distance and assuming their GW emission are all detected, we ensure that the only parameters which determine an merger's status as a standard siren are its inclination angle and $H_0$. In this way, we probe EM anisotropy bias in the absence of GW anisotropy bias, and can thus investigate the degree to which our method uniquely corrects for EM anisotropy bias.

In determining the value of the bias, we must compare the mean of the inferred $H_0$ posterior to some true value. While in strict terms the true value of $H_0$ is that assumed in the generation of the BNS mergers, it is inappropriate to use this as the baseline for the bias since it is not necessarily the value which would be inferred if all 10 BNS mergers were EM+GW mergers. Recall that EM anisotropy bias is the difference between the value of $H_0$ inferred from a sample of EM+GW mergers and the value inferred a subset of those same mergers.

The biased and corrected posteriors produced in this test are shown in Figure \ref{fig:violin_results}. Both the biased and corrected distributions exhibit a long tail to high $H_0$, which is due to the low inclinations of the three EM+GW mergers. Even after our correction scheme is applied, some small probability density is still applied to these high $H_0$. The mean of the biased distribution is $70.17^{+1.31}_{-1.36}$ km s$^{-1}$ Mpc$^{-1}$ while that of the corrected distribution is $70.01^{+1.18}_{-0.75}$ km s$^{-1}$ Mpc$^{-1}$. The corrected distribution accounts for $94.12$\% of EM anisotropy bias. This is greater than the correction level achieved for GW anisotropy bias in both \cite{gerardi2021} and the GW anisotropy-only test presented in this work. The correction itself manifests as a suppression of high $H_0$ values, as is expected from our approach to EM anisotropy bias correction (see Section \ref{sec:emc}).

\begin{table}
    \centering
    \begin{tabular}{|l|l|l|}
    \hline
    \textbf{Test Name} & \textbf{Exposure Depth} ($m_\text{AB}$) & \textbf{Follow-Up Scheme}       \\ \hline
    C1 & $23.3$ & Perfect  \\ \hline
    C2 & $23.3$ & GW190425+GOTO \\ \hline
    C3 & $25$ & Perfect \\ \hline
    C4 & $25$ & GW190425+GOTO \\ \hline
    C5 & $25$ & GW190814+CFHT \\ \hline
    \end{tabular}
    \caption{The five validation tests performed to evaluate our method's ability to correct for both EM and GW anisotropy biases. These tests vary the exposure depth of the assumed EM follow-up campaign as well as the coverage of the campaign. The form of the follow-up schemes listed is (GW localization)+(telescope), except when the follow-up is assumed to be perfect. A `perfect' follow-up campaign assumes that the targeted merger lies within the imaged region of the sky. The $H_0$ posteriors inferred in C1 are shown in Figure \ref{fig:violin_results} and those inferred in C3 and C5 are shown in Figure \ref{fig:violin_results2}.}
    \label{tab:combined_tests}
\end{table}

\subsection{Combined Correction Tests}

We performed five tests to gauge the ability of our method to simultaneously correct for EM and GW anisotropy bias. Each test considers the same sample of mergers used in the GW-only correction test, shown in Figure \ref{fig:100_dl}. In these tests, we vary the EM follow-up campaign specifications for the same $100$ mergers. In two of these tests, the EM follow-up campaign is assumed to have total sky coverage in the $z$ band. The tests differ by the exposure depth assumed, which is $23.3$ $m_\text{AB}$ in one and $25.0$ $m_\text{AB}$ in the other. For each exposure depth, we also test how our method can correct for bias if a specific GW localization and follow-up campaign is assumed. We perform this using the GW localization of GW190425 and its GOTO follow-up campaign \citep{Steeghs_2022}. Each test and its specifications is laid out in Table \ref{tab:combined_tests}.

\subsubsection{C1 and C2}

In C1 and C2, we test the ability of our method to correct for EM and GW anisotropy bias when the detection threshold for EM follow-up is set to $23.3$ $m_\text{AB}$ in the $z$-band. Three posteriors are produced in C1. First, the EM+GW mergers are used to perform a biased inference on $H_0$. Then an intermediate correction is performed by activating the GW anisotropy correction method in the forward model. The final posterior is fully-corrected, accounting for both GW and EM selection effects in the forward model. These posteriors are shown in purple in Figure \ref{fig:violin_results}.

We found that for this sample of EM+GW mergers, the bias was negative, with an inferred $H_0$ of $69.58^{+0.54}_{-0.61}$ km s$^{-1}$ Mpc$^{-1}$. This is not surprising, and it is fairly common when the EM detection threshold is low compared to the magnitudes of BNS mergers at large distances ($\gtrsim 200$ Mpc) where GW anisotropy bias becomes important. \cite{gerardi2021} demonstrates that negatively-biased EM+GW merger posteriors are common in $100$-merger sample. 

Applying GW anisotropy bias correction raises the inferred value of $H_0$ to $69.89^{+0.65}_{-0.47}$ km s$^{-1}$ Mpc$^{-1}$, correcting for $50.00\%$ of the bias. This demonstrates that our method for GW anisotropy bias correction works whether the bias is positive or negative. Including EM anisotropy changes the inferred value of $H_0$ to $69.93^{+0.59}_{-0.51}$ km s$^{-1}$ Mpc$^{-1}$, increasing the level of bias correction to $68.18\%$.

In C2, the GW localization of GW190425 and the GOTO follow-up campaign for that merger is assumed. Due to the poor localization coverage of the follow-up, this resulted in our method being unable to correct for EM anisotropy bias, although the degree of GW anisotropy bias correction remained the same. This is sensible, since any given GW-only merger is far more likely to lie outside the EM follow-up coverage than to be too dim to be seen by the detector. 

\subsubsection{C3 and C4}

C3 and C4 consider the same sample of $100$ mergers as the other combined tests, while assuming that the EM follow-up detection sensitivity reaches $25$ $m_\text{AB}$. C3 assumes that all mergers lie within the follow-up campaign's sky coverage, while C4 assumes the sky localization of GW190425 and that merger's GOTO EM follow-up campaign for all mergers. The posterior distributions for $H_0$ produced in C3 is shown in Figure \ref{fig:violin_results2}, referred to therein as the `unrealistic follow-up' case.

When perfect localization coverage is assumed (C3), the biased posterior of $H_0$ is $69.49^{+0.69}_{-0.51}$ km s$^{-1}$ Mpc$^{-1}$. Once GW anisotropy correction is applied, this becomes $70.47^{+1.50}_{-1.48}$ km s$^{-1}$ Mpc$^{-1}$. GW anisotropy correction in this sample of mergers acts to broaden the posterior significantly due uncertainty in the level of GW anisotropy bias itself. Once EM anisotropy bias is also corrected for, the inferred value of $H_0$ becomes $69.96^{+1.16}_{-1.06}$ km s$^{-1}$ Mpc$^{-1}$, constituting a bias correction level of $92.12\%$ at the cost of broadening the posterior by roughly a factor of $2$.

Assuming realistic localization coverage (C4), the biased posterior of $H_0$ is $70.61^{+1.06}_{-0.90}$ km s$^{-1}$ Mpc$^{-1}$. Applying GW anisotropy bias correction changes this to $70.38^{+1.01}_{-0.96}$, constituting a $37.71\%$ correction in the bias level. Applying the EM anisotropy bias correction in addition to this changes the inferred value of $H_0$ to $69.81^{+0.99}_{-1.09}$, constituting a $68.85\%$ reduction in the bias level. This demonstrates that increasing the depth of follow-up imaging can allow for EM anisotropy bias correction even when the localization coverage is poor.

\subsubsection{C5}

C5 is similar to C4, except that the sky localization of GW190814 is used along with that merger's Canada-France-Hawaii Telescope follow-up campaign. The purpose of this test is to illustrate the quality with which EM and GW anisotropy biases may be corrected for when exposures are deep and the EM follow-up campaign coverage is extensive. In this test, the biased posterior of $H_0$ is $69.40^{+1.22}_{-1.14}$ km s$^{-1}$ Mpc$^{-1}$. Once GW anisotropy correction is applied, this becomes $70.23^{+1.38}_{-0.96}$ km s$^{-1}$ Mpc$^{-1}$. Once EM anisotropy bias is also corrected for, the inferred value of $H_0$ becomes $70.14^{+1.20}_{-1.34}$ km s$^{-1}$ Mpc$^{-1}$, constituting a bias correction level of $76.67\%$. The posterior distributions for $H_0$ produced in this test are displayed in Figure \ref{fig:violin_results2}, labelled therein as the `realistic follow-up' case. Due to the incompleteness of the follow-up campaign, all posteriors are broader than those produced in the case where perfect follow-up is assumed (C3). We find that bias correction in a regime where EM follow-up is nearly complete achieves a similar level of bias correction to C3.

\section{Conclusions}
\label{sec:conc}

In this study, we have demonstrated how simulation-based inference can be used to produce an unbiased measurement of $H_0$ using mergers detected with EM+GW in addition to GW-only mergers. In doing so, we account for both GW and EM selection effects. Our GW anisotropy bias correction method matches the performance of the SBI method presented by \cite{gerardi2021} and further generalizes its inference to account for EM anistropy bias in standard siren measurements of $H_0$.

The key to EM anisotropy correction is the inclusion of GW-only events -- mergers whose GW signal is detected in the absence of a detected EM counterpart. For many such mergers, the fact of EM non-detection places a constraint on the merger's apparent magnitude, which in turn constrains the range of possible inclination angles and luminosity distances for the merger. These constraints in turn act to temper, and in many cases fully remove, EM anisotropy bias in the inferred value of $H_0$.

We found that the inclusion of EM anisotropy correction in scenarios where EM anistropy bias is negligible can reduce the bias correction level of our method by a few percent. However, this slight reduction in correction efficacy in low-EM anisotropy bias scenarios is outweighed by the high efficacy of EM anisotropy bias correction in high-EM anisotropy bias scenarios. Without an ab initio method of determining whether a population of BNS mergers is significantly EM anisotropy biased, it is safest to include EM anisotropy correction in the analysis of a merger sample alongside GW anisotropy correction.

The tests which consider the GW footprint of GW190425 and its GOTO EM follow-up campaign (C3 and C4) illustrate the importance of a comprehensive EM follow-up campaign for each merger. In these tests, the probability of a merger not lying within the EM follow-up region is considered. When the probability that a GW-only merger only lacks a detected EM counterpart because it was not localized within any telescope pointings is high, only weak constraints can be placed on that merger's inclination angle and luminosity distance. Such was the case for GW190425, where the localization was poor and the GOTO campaign only had $\sim$25\% localization coverage. This in turn weakens the correction level on the EM anisotropy bias. Thus, a maximally-informative EM follow-up campaign should seek to maximize its coverage of the BNS merger's sky localization. We make this point by including a test where we consider a more complete EM follow-up scenario, the Canada-France-Hawaii Telescope follow-up of GW190814 (C5), wherein the bias correction level is comparable to that of perfect localization coverage.

Future extensions of this work should focus on more realistic inference scenarios. For example, in this work we assumed the neutron star masses of each merger to be precisely measured from the GWs, when in reality a GW measurement provides a joint probability distribution for the merger's neutron star masses. We also assumed the same EM follow-up routine for each merger with the same underlying sky localization. A more realistic treatment of EM follow-up should consider the real sky localizations of each merger with merger-specific EM follow-up routines. Future work should also address the uncertainty in the kilonova parameters themselves, as incorrectly assuming their values can lead to either insufficient or overzealous bias correction.

The significance of the biases addressed in this study will only increase as more BNS mergers are detected in the coming years, underscoring the importance of a reliable bias correction method. SBI-based approaches such as ours enable cosmologists to take full advantage of the incoming deluge of BNS merger detections to produce unbiased measurements of the Hubble constant, and possibly resolve the Hubble tension.

\section*{Acknowledgments}

The authors thank Sabrina Berger, Michael Matesic, Carter Rhea, Jason Rowe, Nicholas Vieira, and Clovis Vinant-Tang for helpful discussions. S.G.H. acknowledges support from the Natural Sciences and Engineering Research Council of Canada (NSERC) through their Canada Graduate Scholarships - Master's programme, as well as from the Bishop's University Foundation through their Graduate Entrance Scholarship. J.J.R.\ and D.H.\ acknowledge support from the Canada Research Chairs (CRC) program, the NSERC Discovery Grant program, the FRQNT Nouveaux Chercheurs Grant program, and the Canadian Institute for Advanced Research (CIFAR). J.J.R. acknowledges support from \ the Canada Foundation for Innovation (CFI), and the Qu\'{e}bec Ministère de l’\'{E}conomie et de l’Innovation. Computations were performed on the Cedar and B\'{e}luga supercomputing clusters managed by Compute Canada.

\section*{Data Availability}
The data underlying this article is available upon request. The inference code is available on the author’s GitHub page: \url{https://github.com/samgagnon/kilonova_sim}.




\bibliographystyle{mnras}
\bibliography{example} 



\appendix

\section{Methods Supplemental Material}

\subsection{Kilonova Parameters}
\label{apdx:params}

We generate kilonova light curves using the \texttt{MOSFiT} software. We assume that kilonovae have two merger ejecta components, one red and one blue \citep{kn_review}, with opacities $\kappa_\text{red}$ and $\kappa_\text{blue}$ respectively. We use fiducial values of $\kappa_{\text{red}}=10$ cm$^2$ g$^{-1}$ and $\kappa_{\text{blue}}=0.5$ cm$^2$ g$^{-1}$, following the results of \cite{ejecta2018}.

A kilonova's peak luminosity scales with the quantity of ejecta produced in the NS merger, and thus to the masses of the NS themselves. Our light curve simulation requires as inputs the BNS merger chirp mass, 

\begin{equation}
    \mathcal{M}=\left(\frac{(m_1m_2)^{3/5}}{(m_1+m_2)^{1/5}}\right),
\end{equation}

\noindent and NS mass ratio, $q=m_1/m_2$, where $m_1<m_2$. We compute these quantities from the true NS masses provided by \texttt{GWToolbox}.

The luminosities of red and blue components also depend on their temperatures, which are influenced by the disk ejection fraction, $\epsilon_\text{disk}$, and the blue component enhancement factor $\alpha$. $\epsilon_{\text{disk}}$ governs the fraction of the BNS remnant accretion disk ejected post-merger. We use a fiducial value of $\epsilon_\text{disk}=0.15$, while \cite{nicholl2021} state that the true value may range anywhere from $0.05$ to $0.5$, following \cite{kn_review}. Meanwhile, surface winds from the merger remnant enhance the temperature of the blue ejecta, encapsulated by $\alpha$. \cite{nicholl2021} assign $\alpha$ a flat prior from $0.1$ to $1.0$, while we set it to $1.0$ (maximum enhancement).

Finally, the geometry of the kilonova influences which component we actually observe. In \texttt{MOSFiT}, kilonova geometry is simplified into two distinct angular regions defined by the half-opening angle of the blue component, $\theta$. We follow \cite{nicholl2021} by setting $\theta=45^\circ$. BNS mergers also produce an associated gamma-ray burst (GRB), which shocks ejecta material in its vicinity. For the purposes of this study, we assume that the GRB shock negligibly effects the blue component, thus setting the half-opening angle of the GRB shock to $\theta_c=0$. We do this since the half-opening angle of the shocked cone is uncertain and its physics are poorly understood.

\subsection{Summary Statistic}
\label{apdx:sumstat}

We measure the consistency of $\Theta$ with $\mathbf{x}$ explicitly using the $(D_L,\iota)$ distributions of each GW-detected merger as measured from its gravitational wave emission. The likelihood of guessed parameters $D_L$ and $\iota$ for any particular merger are taken as $\mathcal{L}=P(D_L,\iota|x_{\text{GW}})$ where $x_{\text{GW}}$ is the gravitational wave strain for that merger. Each merger's likelihood contributes to the informative summary statistic $\mathcal{S}$, defined as

\begin{equation}
    \mathcal{S}=\sum_i^N \log(\mathcal{L}(D_{L,i},\iota_i|x_{\text{GW},i})),
    \label{eq:sumstat}
\end{equation}

\noindent or the sum of log-likelihoods of each merger. We use this definition of $\mathcal{S}$ since it is based on an analytic likelihood, allowing the neural ratio estimator to easily learn the true likelihood-to-evidence ratio. Furthermore, variance in $\mathcal{S}$ originates entirely from uncertainty in the proposed parameters $D_{L,i}$ and $\iota_i$, meaning that our selected summary statistic does not introduce any new biases that are not already present in the source data. This follows the principles of summary statistic selection laid out by \cite{sumstat}.

Our choice of the minimum value of $\mathcal{S}$ for null returns follows from the form of Equation \ref{eq:sumstat}. Since the neural ratio estimator is trained to maximize the log-likelihood of the target parameter, a null return must be lower than the lowest informative return. To this end, we set a minimum allowable likelihood for each merger within a sample, $\mathcal{L}_{\text{min}}=10^{-6}$. We choose this as the minimum likelihood since it is equivalent to the likelihood of sampling $D_L$ and $\iota$ from a 2D uniform distribution, given that our likelihood distributions have a resolution of $1000\times1000$. As such, any sampled $(D_L,\iota)$ with this likelihood or below is no more informative than a sample drawn from a uniform distribution, and we therefore deem it `non-informative'. If a parameter sample is non-informative for all mergers in a data set (i.e., if $\mathcal{S}<-6\cdot N$) then the summary statistic is automatically returned as null, or $\mathcal{S}_{\text{null}}=-6\cdot N$.

A careful inspection of Equation \ref{eq:sumstat} reveals that its maximum and minimum vary with the number of mergers in a sample. Furthermore, only a fraction of all BNS mergers within a sample have detected GWs in any given sample. To maintain consistent bounds for the summary statistic, we treat each merger not detected in GWs as contributing their maximum log-likelihood to the summary statistic. Thus, the summary statistic for each sample is computed as

\begin{equation}
    \mathcal{S}=\sum_i^D l_i(\Theta | D_{L,i},\iota_i)+\sum_i^N \text{max}(l_i(\Theta)),
\end{equation}

\noindent where $D$ is the number of GW-detected mergers, $N$ is the number of mergers not detected in GWs, $l_i(\Theta)$ is the log-likelihood of an merger indexed $i$ with the parameter sample $\Theta$, and $(D_{L,i},\iota_i)$ are the luminosity distance and inclination angle sampled for the merger $i$.

\subsection{Intermediate Processing}
\label{apdx:ip}

Producing an unbiased $H_0$ posterior from $\mathcal{S}_{\textrm{GW}}$ is not as simple as training another MNRE. This is because the nonzero returns in $\mathcal{S}_{\textrm{GW}}$ are noise-dominated. Each summary statistic $\mathcal{S}$ is the sum of log-likelihoods for each merger given the sampled parameter vector. EM+GW mergers are governed by only two free variables: $H_0$ and $\iota$, the former of the two is shared for all mergers within a sample. This means that at a particular $H_0$, the variance in log-likelihoods is entirely determined by the uncertainty in $\iota$ for each merger. The redshift of a EM+GW merger is known from EM observations, fixing the luminosity distance when a cosmology is assumed ($(H_0',z)\xrightarrow[]{}D_L'$). Meanwhile, GW-only mergers are governed by an additional parameter, $D_L$. Without a known redshift, the luminosity distance of a GW-only merger is free to be sampled independently of $H_0$. This means that for a particular $H_0$, the variance in the log-likelihood is driven by both the uncertainty in $D_L$ and the uncertainty in $\iota$. We find that this drives the variance in nonzero returns to a point where an MNRE is no longer capable of recovering the underlying pattern, rendering direct posterior inference impossible.

With nonzero returns rendered useless, we are forced to look to the zero returns for answers. Zero returns are not non-informative: they show the values of $H_0$ which are forbidden given a sample of mergers. Taking the ratio between the number of zero returns and the number of total samples as a function of $H_0$, the `rejection' rate, or $R(H_0)$, may be constructed. Any set of summary statistics has an associated rejection rate function, and we denote the rejection rate produced from $\mathcal{S}_{\textrm{S}}$ as $R_{\textrm{S}}(H_0)$, and that produced from $\mathcal{S}_{\textrm{GW}}$ as $R_{\textrm{GW}}(H_0)$.

These rejection rates differ from one another, as including GW-only mergers increases the selectiveness of the forward model. Crucially, the selectiveness does not change uniformly with $H_0$, but follows some functional relationship encoded within $R_{\textrm{GW}}(H_0)$. The objective in correcting for EM anisotropy bias is thus to modify the set $\mathcal{S}_{\textrm{S}}$ such that it accounts for the information in $R_{\textrm{GW}}(H_0)$. A responsible modification to $\mathcal{S}_{\textrm{S}}$ is one that does not introduce any information beyond what is encoded in GW-only mergers. The most obvious modification is to apply $R_{\textrm{GW}}(H_0)$ to $\mathcal{S}_{\textrm{S}}$ by changing some portion of the informative summary statistics to zero returns according to the target rejection rate. This does not change the posterior learned from the distribution since the MNRE explicitly ignores null returns in favour of informative samples. Therefore, the solution must modify the informative samples in $\mathcal{S}_{\textrm{S}}$ without setting them to zero. We define the modified set of summary statistics as the product of $\mathcal{S}_{\textrm{S}}$ and $R_{\textrm{GW}}(H_0)$, i.e.

\begin{equation}
    \mathcal{S}_{\textrm{S}}'=\mathcal{S}_{\textrm{S}}|\mathcal{S}_{\textrm{GW}}=R_{\textrm{GW}}(H_0)\cdot \mathcal{S}_{\textrm{S}}.
\end{equation}

This modified set of summary statistics, $\mathcal{S}_{\textrm{S}}'$, includes all information granted by the EM+GW mergers, while also incorporating the rejection rate required by GW-only mergers. $H_0$ posteriors learned from this modified set correct for the bias introduced by EM anisotropy without broadening the distribution by including noisy GW-only mergers.

\subsection{Neural Ratio Estimation}
\label{apdx:nre}

Neural ratio estimation (NRE) involves training a classifier network $d_\phi:\Theta\times X\xrightarrow[]{}[0,1]$ to discriminate between pairs $(\theta,x)$ sampled from the joint posterior distribution $p(\theta,x)$ and the product of the marginals $p(\theta)p(x)$. By the Neyman-Pearson lemma, the probability that a data pair belongs to the joint posterior is proportional to the posterior density \citep{pearsonlemma}. Formally, the neural network optimizes

\begin{equation}
    \phi^*=\arg\min_\phi \underset{p(\theta,x)p(\theta')}{\mathbb{E}}[\mathcal{L}(d_\phi(\theta,x))+\mathcal{L}(1-d_\phi(\theta',x))],
\end{equation}

\noindent where $\mathcal{L}(p)=-\log p$ is the negative log-likelihood. For this task, the Bayes optimal classifier uses the decision function

\begin{equation}
    d(\theta,x)=\frac{p(\theta,x)}{p(\theta,x)+p(\theta)p(x)},
\end{equation}

\noindent which defines the likelihood-to-evidence (LTE) ratio

\begin{equation}
    r(\theta,x)=\frac{d(\theta,x)}{1-d(\theta,x)}=\frac{p(\theta,x)}{p(\theta)p(x)}=\frac{p(x|\theta)}{p(x)}=\frac{p(\theta|x)}{p(\theta)}.
\end{equation}

\noindent Thus, NRE grants us an estimator $\log r_\phi(\theta,x)=\text{logit}(d_\phi(\theta,x))$ of the LTE log-ratio and a surrogate $\hat{p}(\theta|x)=r_\phi(\theta,x)p(\theta)$ for the posterior density.


\bsp	
\label{lastpage}
\end{document}